\documentclass[11pt,preprint]{aastex}

\slugcomment{Accepted for publication in the Astrophysical Journal}

\shorttitle{Ly$\alpha$-He~{\sc ii} Dual Emitters}
\shortauthors{Nagao et al.}

\begin{document}

\title{A photometric survey for Ly$\alpha$-He~{\sc ii} dual emitters:\\
       Searching for Population III stars in high-redshift galaxies}

\author{
          Tohru Nagao            \altaffilmark{1},
          Shunji S. Sasaki       \altaffilmark{2,3},
          Roberto Maiolino       \altaffilmark{4},
          Celestine Grady        \altaffilmark{5},\\
          Nobunari Kashikawa     \altaffilmark{1},
          Chun Ly                \altaffilmark{5},
          Matthew A. Malkan      \altaffilmark{5},
          Kentaro Motohara       \altaffilmark{6},\\
          Takashi Murayama       \altaffilmark{2},
          Daniel Schaerer        \altaffilmark{7},
          Yasuhiro Shioya        \altaffilmark{3,8},
          Yoshiaki Taniguchi     \altaffilmark{3,8}
}

\altaffiltext{1}{
    National Astronomical Observatory of Japan,
    2-21-1 Osawa, Mitaka, Tokyo 181-8588, Japan
}
\altaffiltext{2}{
    Astronomical Institute, Graduate School of Science,
    Tohoku University, Aramaki, Aoba, Sendai 980-8578, Japan
}
\altaffiltext{3}{
    Department of Physics, Graduate School of Science and Engineering,
    Ehime University, 2-5 Bunkyo-cho, Matsuyama 790-8577, Japan
}
\altaffiltext{4}{
    INAF -- Osservatorio Astrofisico di Roma,
    Via di Frascati 33, 00040 Monte Porzio Catone, Italy
}
\altaffiltext{5}{
    Department of Physics and Astronomy, 
    University of California at Los Angeles, 
    P. O. Box 951547, Los Angeles, CA 90095-1547, USA
}
\altaffiltext{6}{
    Institute of Astronomy, Graduate School of Science, 
    University of Tokyo, 2-21-1 Osawa, Mitaka, Tokyo 181-0015, Japan
}
\altaffiltext{7}{
    Geneva Observatory, University of Geneva, 
    51 chemin des Maillettes, 1290 Sauverny, Switzerland 
}
\altaffiltext{8}{
    Research Center for Space and Cosmic Evolution,
    Ehime University, 2-5 Bunkyo-cho, Matsuyama 790-8577, Japan
}

\begin{abstract}
  We present a new photometric search for high-$z$ galaxies
  hosting Population III (PopIII) stars based on deep 
  intermediate-band imaging observations obtained in the 
  Subaru Deep Field (SDF), by using Suprime-Cam on the 
  Subaru Telescope. By combining our new data with the existing 
  broad-band and narrow-band data, we searched for galaxies 
  which emit strongly both in Ly$\alpha$ and in 
  He~{\sc ii}$\lambda$1640 (``dual emitters'') that are 
  promising candidates for PopIII-hosting galaxies, at 
  $3.93 \la z \la 4.01$ and $4.57 \la z \la 4.65$. Although we 
  found 10 ``dual emitters'', most of them turn out to be 
  [O~{\sc ii}]-[O~{\sc iii}] dual emitters or 
  H$\beta$-(H$\alpha$+[N~{\sc ii}]) dual emitters at $z < 1$, 
  as inferred from their broad-band colors and from the ratio 
  of the equivalent widths. No convincing candidate of
  Ly$\alpha$-He~{\sc ii} dual emitter of
  $SFR_{\rm PopIII} \ga 2 M_\odot$ yr$^{-1}$ was found by our 
  photometric search in $4.03 \times 10^5$ Mpc$^3$ in the SDF. 
  This result disfavors low feedback models for PopIII star 
  clusters, and implies an upper-limit of the PopIII SFR density
  of $SFRD_{\rm PopIII} < 5 \times 10^{-6} M_\odot$ yr$^{-1}$
  Mpc$^{-3}$. This new selection method to search for 
  PopIII-hosting galaxies should be useful in future 
  narrow-band surveys to achieve the first observational 
  detection of PopIII-hosting galaxies at high redshifts.
\end{abstract}


\section{Introduction}

Population III (PopIII) stars are those formed out of 
primordial gas, enriched only through Big-Bang 
nucleosynthesis. Since massive PopIII stars are promising 
candidates as sources for cosmic reionization (e.g., 
Ciardi et al. 2000; Loeb \& Barkana 2001; Wyithe \& Loeb 2003; 
Sokasian et al. 2004) and an important population for early 
phases of the cosmic chemical evolution (e.g., 
Wasserburg \& Qian 2000; Abia et al. 2001; 
Qian \& Wasserburg 2001; Bromm et al. 2003), their properties 
have been extensively investigated from the theoretical point 
of view. PopIII stars have not been discovered yet; obviously, 
their direct detection and the observational studies of their 
properties would provide a completely new and important step 
toward understanding the evolution of galaxies. The expected
observables of high-$z$ galaxies hosting PopIII stars have 
been theoretically investigated in recent years. Such 
galaxies are expected to show strong Ly$\alpha$ emission, 
with an extremely large equivalent width (EW), and moderately 
strong He~{\sc ii}$\lambda$1640 emission (e.g., 
Tumlinson \& Shull 2000; Tumlinson et al. 2001; 
Oh et al. 2001; Schaerer 2002, 2003; Tumlinson et al. 2003),
due to the high effective temperature up to $\sim10^5$ K of 
PopIII stars (e.g., Bromm et al. 2001; Tumlinson et al. 2003).

Most models predict that PopIII stars dominated the
re-ionization of the universe at $7 \la z \la 15$. However,
they also predict that PopIII stars may still exist at 
redshifts currently accessible with $8-10$m-class 
telescopes, i.e., $z<7$, although it may depend both on some 
model parameters of PopIII [e.g., initial mass function (IMF)] 
and on some environmental parameters such as the mixing 
efficiency (e.g., Scannapieco et al. 2003, 2006; Jimenez \& 
Haiman 2006; Schneider et al. 2006; Brook et al. 2007; 
Tornatore et al. 2007). Some observations have found 
Ly$\alpha$ emitters (LAEs) at $z > 4$ with a very large EW,
which is hard to explain through star-formation without 
PopIII (e.g., Malhotra \& Rhoads 2002; Nagao et al. 2004, 
2005a, 2007; Shimasaku et al. 2006; 
Dijkstra \& Wyithe 2006). However, the search for 
He~{\sc ii}$\lambda$1640 emission as direct evidence for the 
presence of PopIII in such galaxies is far more controversial. 
Jimenez \& Haiman (2006) pointed out the possible 
He~{\sc ii}$\lambda$1640 signature in the composite spectrum 
of $\sim$1000 Lyman-break galaxies (LBGs) at $z \sim 3$ made 
by Shapley et al. (2003), although the 
He~{\sc ii}$\lambda$1640 feature in the LBG composite 
spectrum may be attributed to a stellar wind feature 
associated with massive stars as mentioned by Shapley et al.
(2003). On the other hand, other searches for 
He~{\sc ii}$\lambda$1640 in higher-$z$ galaxies have failed, 
through stacking analysis of LAEs (Dawson et al. 2004; 
Ouchi et al. 2008) or through ultra-deep near-infrared 
spectroscopy of an individual LAE (Nagao et al. 2005b).

Nevertheless, the He~{\sc ii}$\lambda$1640 emission from 
PopIII-hosting galaxies may already be detected in current 
deep narrow-band (NB) surveys (mostly aiming for LAE 
searches) as NB-excess objects, but not identified as
He~{\sc ii} emitters (Tumlinson et al. 2001) since NB surveys 
are more sensitive to faint emission lines than spectroscopic
observations. Galaxies in a young PopIII-hosting phase are 
expected to show He~{\sc ii}$\lambda$1640 emission with 
$EW_{\rm rest} > 20{\rm \AA}$ (e.g., Schaerer 2003), which 
corresponds to a redshifted emission line at 
$\sim$9200${\rm \AA}$ with $EW_{\rm obs} > 110{\rm \AA}$ for 
galaxies at $z \sim 4.6$. Such He~{\sc ii}-emitting galaxies 
could already be present as NB-excess objects in current 
deep NB surveys for LAEs at $z \sim 6.5$ (see, e.g., 
Taniguchi et al. 2005a; Kashikawa et al. 2006). If a 
NB-excess object is due to He~{\sc ii}$\lambda$1640 emission, 
then the same object should show stronger Ly$\alpha$ 
emission at a shorter wavelength, since the PopIII-hosting 
galaxies should emit Ly$\alpha$ with 
$EW_{\rm rest} > 500{\rm \AA}$ (e.g., Schaerer 2003). 
Therefore, by performing additional NB (or intermediate-band) 
imaging observations whose wavelength is matched to the 
redshifted Ly$\alpha$, we may be able to find 
``Ly$\alpha$-He~{\sc ii} dual emitters'' that are promising 
candidates for PopIII-hosting galaxies. Motivated by these 
considerations, we performed new intermediate-band imaging 
observations on the Subaru Deep Field (SDF; Kashikawa et al. 
2004), where very deep and wide broad-band and NB imaging 
data are available\footnote{
   The reduced data and the object catalogs are available
   at {\tt http://soaps.naoj.org/}.}.

In this paper, we report new intermediate-band imaging 
observations for the SDF, describe the combination of the new
data with existing data to search for Ly$\alpha$-He~{\sc ii} 
dual emitters at $z \sim$4.0 and $\sim$4.6, and discuss the
inferred constraints on the population of PopIII-hosting
galaxies. Throughout this paper, we adopt a cosmology with 
($\Omega_{\rm tot}, \Omega_{\rm M}, \Omega_{\rm \Lambda}$)
= (1.0, 0.3, 0.7) and $H_0 = 70$ km s$^{-1}$ Mpc$^{-1}$.
We use the AB photometric system for optical magnitudes. 

\section{Data}

\subsection{Method and filter selection}

The field investigated in this project is the SDF, centered 
at $\alpha$(J2000) = 13:24:38.9 and $\delta$(J2000) =
+27:29:25.9 (Kashikawa et al. 2004), where the Galactic
dust extinction is low ($E_{B-V} = 0.017$ mag; 
Schlegel et al. 1998). The optical photometric data obtained 
in the SDF so far are summarized in Table 1. Among the 5 
existing NB images, we focus on the NB816 and NB921 data 
to search for the putative He~{\sc ii}$\lambda$1640 emission 
from PopIII-hosting galaxies. Note that the central 
wavelengths and the half-widths of the transmittance of 
NB816 and NB921 are ($\lambda_{\rm c}$, 
$\Delta\lambda_{\rm FWHM}$) = (8150${\rm \AA}$, 
120${\rm \AA}$) and (9196${\rm \AA}$, 132${\rm \AA}$), 
respectively. We did not consider the NB973 data, which are
too shallow to search for He~{\sc ii}$\lambda$1640 emission 
from PopIII-hosting galaxies. We also did not use the NB704 
and NB711 data, because their wavelengths are too blue,
resulting in too low redshifts ($z \sim 3.3$ for 
He~{\sc ii}$\lambda$1640 emitters). The NB816 and NB921 
filters can be used to search for He~{\sc ii}$\lambda$1640 
emitters at $3.93 \la z \la 4.01$ or $4.57 \la z \la 4.65$, 
respectively. If there are He~{\sc ii}$\lambda$1640 emitters 
in these redshift ranges, they should show very strong 
Ly$\alpha$ emission at 
$5992{\rm \AA} \la \lambda_{\rm obs} \la 6089{\rm \AA}$ or
$6769{\rm \AA} \la \lambda_{\rm obs} \la 6867{\rm \AA}$.
We then used two intermediate-passband filters (``IA filter 
system''; see, e.g., Yamada et al. 2005; 
Taniguchi et al. 2005b), whose wavelengths correspond to 
these Ly$\alpha$ observed wavelengths with broader 
transmission FWHM than NB filters. Specifically, we used 
IA598 and IA679, whose central wavelengths and half-widths 
of transmittance are 
($\lambda_{\rm c}$, $\Delta\lambda_{\rm FWHM}$) = 
(6008${\rm \AA}$, 298${\rm \AA}$) and (6782${\rm \AA}$, 
339${\rm \AA}$), respectively. Note that the wide IA 
filters select LAEs with a very large EW (e.g., 
Fujita et al. 2003a; Ajiki et al. 2004; Yamada et al. 2005).
This is not a problem for this project, since 
PopIII-hosting galaxies are expected to show Ly$\alpha$ 
emission with a very large EW 
($EW_{\rm rest} > 100{\rm \AA}$; e.g., Schaerer 2002, 2003;
Scannapieco et al. 2003).

In summary, we can select ``Ly$\alpha$-He {\sc ii} dual
emitters'' at $3.93 \la z \la 4.01$ by combining the IA598 
and NB816 data, and at $4.57 \la z \la 4.65$ by combining 
the IA679 and NB921 data. In Figure 1, a schematic view of 
this selection method is shown. The target redshift ranges 
are shown more clearly in Figure 2, where the IA and NB 
filter transmission curves are shown as functions of the 
targeted redshift.

\subsection{Observations and the data reduction}

The SDF was observed on 22 April 2007 (UT) with 
Suprime-Cam [that has a field-of-view (FOV) of $34 \times 27$ 
arcmin$^2$; Miyazaki et al. 2002] on the Subaru Telescope 
(Iye et al. 2004). We used two intermediate-passband filters, 
IA598 and IA679, as described above. The individual exposure 
time was 120 seconds or 900 seconds. A small dithering was 
performed between individual exposures to cover the gaps 
between the detectors and the bad pixels. The typical seeing 
during the observation was 0.6 -- 0.9 arcsec in FWHM. We 
discarded frames with bad seeing (two 900 seconds frames for 
each filter); as a consequence, the total integration times 
used to construct the final combined images of the IA598 and 
IA679 data are 111 minutes and 231 minutes, respectively 
(Table 1). We also observed spectrophotometric standard stars, 
G93-48, GD108, HZ21, and HZ44 for the flux calibration.

The individual CCD data were reduced and combined by using 
the data reduction package SDFRED (Yagi et al. 2002; 
Ouchi et al. 2004). Since the PSF sizes of the reduced and 
combined IA598 and IA679 data (0.91 arcsec and 0.93 arcsec
in FWHM, respectively) are smaller than that of the existing 
SDF imaging data (0.99 arcsec), we matched the PSF of the 
IA598 and IA679 data to the existing SDF data by smoothing
with proper Gaussian kernels. After masking the low-quality 
regions, such as the edge of the FOV and the regions affected 
by bright stars, the effective surveyed area is 875 arcmin$^2$. 
Consequently, the co-moving survey volume at 
$3.93 \la z \la 4.01$ is $2.08 \times 10^5$ Mpc$^3$ and 
$1.95 \times 10^5$ Mpc$^3$ at $4.57 \la z \la 4.65$; the 
total volume surveyed by this study is thus 
$4.03 \times 10^5$ Mpc$^3$.
% 1''(z=3.97)=6.972 kpc, 1''(z=4.61)=6.530 koc
% D_com(z=3.93)=7119.5 Mpc, D_com(z=4.01)=7175.4 Mpc
% D_com(z=4.57)=7532.8 Mpc, D_com(z=4.65)=7579.5 Mpc
% V1 = 875 x 60^2 x 0.006972^2 x (1+3.93)^2 x (7175.4-7119.5) 
%    = 2.08e5 Mpc^3
% V2 = 875 x 60^2 x 0.006530^2 x (1+4.57)^2 x (7579.5-7532.8)
%    = 1.95e5 Mpc^3

\subsection{Source detection and photometry}

By using the final images we extracted the IA598-selected 
and IA679-selected object catalogs. Source detection and 
photometry were performed by using SExtractor version 2.3.2 
(Bertin \& Arnouts 1996). We adopted criteria of 5 
adjacent pixels and 2$\sigma$ above the noise level for 
source detection. Photometry was performed with a 2 arcsec 
diameter aperture for each band image. The 3$\sigma$ limiting 
magnitudes of IA598 and IA679 for 2 arcsec apertures are 
26.52 mag and 27.07 mag, respectively (Table 1). In the 
following sections we used the inferred photometric catalogs 
with a correction for the Galactic reddening of 
$E_{B-V} = 0.017$ mag (Schlegel et al. 1998), adopting the 
extinction curve by Cardelli et al. (1989).

\section{Results}

\subsection{Selection of IA-excess objects}

To select objects with excess emission in the IA filters 
(``IA-excess'' objects), we determined the matched continuum 
magnitudes for IA598 and IA679, which will be identified as 
C598 and C679, hereafter. They are obtained by
$f_{\rm C598} = 0.45 f_{\rm V} + 0.55 f_{\rm R_C}$ and 
$f_{\rm C679} = 0.75 f_{\rm R_C} + 0.25 f_{\rm i^\prime}$
respectively, where $f_{\rm X}$ is the flux density at band X. 
The weighting factors used here are calculated by taking the
effective wavelengths of the related filters into account.
The 3$\sigma$ limiting magnitudes of C598 and C679 are 
estimated to be 27.77 mag and 27.70 mag respectively, without 
correction for Galactic extinction. We then selected 
IA598-excess objects as those matching $all$ of the following 
criteria:
\begin{equation} 
   21.0 < IA598 < 26.52 \ \ [ \ = 3\sigma \ ], 
\end{equation}
\begin{equation} 
   R_{\rm C} > 28.24 \ \ [ \ = 2\sigma \ ],
\end{equation}
\begin{equation} 
   i^\prime > 27.87 \ \ [ \ = 2\sigma \ ],
\end{equation}
\begin{equation} 
   C598 - IA598 > 0.3,
\end{equation}
\begin{equation} 
   C598 - IA598 > 3\sigma(C598 - IA598),
\end{equation}
and similarly IA679-excess objects were selected as those 
matching $all$ of the following criteria:
\begin{equation} 
   21.0 < IA679 < 27.07 \ \ [ \ = 3\sigma \ ], 
\end{equation}
\begin{equation} 
   R_{\rm C} > 28.24 \ \ [ \ = 2\sigma \ ],
\end{equation}
\begin{equation} 
   i^\prime > 27.87 \ \ [ \ = 2\sigma \ ],
\end{equation}
\begin{equation} 
   C679 - IA679 > 0.3,
\end{equation}
\begin{equation} 
   C679 - IA679 > 3\sigma(C679 - IA679).
\end{equation}
The bright-end limit of the IA magnitudes for the 
selection was introduced to avoid saturation and/or 
non-linearity effects. The IA-excess magnitude adopted here 
(0.3 mag) corresponds to a selection limit in terms of EW 
of $EW_{\rm obs} \ga 114 {\rm \AA}$ for IA598, and 
$EW_{\rm obs} \ga 145 {\rm \AA}$ for IA679.

Note that the adopted limiting EWs are much lower than 
intrinsic EWs theoretically expected for PopIII-hosting 
galaxies [$EW_{\rm rest}$(Ly$\alpha$) $\ga$ 100${\rm \AA}$, 
which corresponds to $EW_{\rm obs}$(Ly$\alpha$) $\ga$ 
500${\rm \AA}$]. However, even in low dust-abundance
environments, Ly$\alpha$ photons from H~{\sc ii} regions 
could suffer from resonance scattering through the neutral 
hydrogen and accordingly the surface brightness of the 
Ly$\alpha$ emission could be diminished, resulting in 
lower observed EWs. Taking this possibility into account, 
we set the limiting IA-excess magnitude (i.e., the 
limiting EWs) as described above.

In Figures 3 and 4, we show the selected IA598-excess and
IA679-excess objects on the diagram of $C598-IA598$ vs. 
$IA598$ and that of $C679-IA679$ vs. $IA679$. The numbers 
of selected IA598-excess objects and of IA679-excess 
objects are 133 and 234, respectively.

\subsection{Selection of IA-NB dual emitters}

The IA-excess selected samples also include low-$z$ objects, 
not only LAEs we are focusing on. More specifically, the 
IA598-excess objects may contain [O~{\sc iii}] emitters at 
$z \sim 0.20$, H$\beta$ emitters at $z \sim 0.24$ and 
[O~{\sc ii}] emitters at $z \sim 0.61$. IA679-excess objects 
may contain H$\alpha$ (+ [N~{\sc ii}]) emitters at 
$z \sim 0.03$, [O~{\sc iii}] emitters at $z \sim 0.35$,
H$\beta$ emitters at $z \sim 0.40$ and [O~{\sc ii}] emitters
at $z \sim 0.82$. Note that fainter emission lines, such as 
[Ne~{\sc iii}] and higher-order Balmer lines, are expected to 
be excluded from the IA-excess object samples, since we are 
sensitive only to relatively high-EW objects 
($EW_{\rm obs} \ga 100 {\rm \AA}$).

Generally, such low-$z$ interlopers should be removed before 
analyzing any statistical properties of high-$z$ LAEs (see, 
e.g., Ajiki et al. 2003). However, we are now focusing on 
``IA-NB dual emitters'' as candidates for 
``Ly$\alpha$-He~{\sc ii} dual emitters'' (i.e., 
PopIII-hosting galaxies). Since low-$z$ H$\alpha$ and 
[O~{\sc iii}] emitters do not show any emission-line features 
around corresponding NB wavelengths 
($\sim 8150 \pm 60 {\rm \AA}$ for IA598 emitters and 
$\sim 9196 \pm 66 {\rm \AA}$ for IA679 emitters), IA598-NB816 
dual emitters and IA679-NB921 dual emitters do not contain 
them. The possible low-$z$ contamination in IA-NB dual 
emitter samples is from [O~{\sc ii}]-[O~{\sc iii}] dual 
emitters and H$\beta$-(H$\alpha$+[N~{\sc ii}]) dual emitters, 
because the wavelength ratios of Ly$\alpha$/He~{\sc ii},
[O~{\sc ii}]/[O~{\sc iii}], and H$\beta$/H$\alpha$ are so 
similar ($\sim$0.741, $\sim$0.744, and $\sim$0.741, 
respectively). More specifically, [O~{\sc iii}] emitters 
observed as NB816 and NB921 emitters are at 
$0.62 \la z \la 0.64$ and $0.82 \la z \la 0.85$, which show 
[O~{\sc ii}] emission at 
$6020{\rm \AA} \la z \la 6110{\rm \AA}$ and 
$6800{\rm \AA} \la z \la 6890{\rm \AA}$ that are covered by
the IA598 and IA679 filters respectively, as shown in 
Figure 5. Similarly, H$\alpha$ (+ [N~{\sc ii}]) emitters 
observed as NB816 and NB921 emitters are at 
$0.23 \la z \la 0.25$ and $0.39 \la z \la 0.41$, which show 
H$\beta$ emission at 
$5990{\rm \AA} \la z \la 6080{\rm \AA}$ and 
$6760{\rm \AA} \la z \la 6860{\rm \AA}$ that are covered by
the IA598 and IA679 filters respectively, as shown in 
Figure 5. Therefore, we select IA-NB dual emitters at first, 
and then will classify them into Ly$\alpha$-He {\sc ii},
[O~{\sc ii}]-[O~{\sc iii}], and 
H$\beta$-(H$\alpha$+[N~{\sc ii}]) dual emitters.

In Figures 6 and 7, we investigate possible NB816 excesses 
of IA598-excess objects in the diagram of $C816-NB816$ vs.
$NB816$, and possible NB921 excesses of IA679-excess 
objects in the diagram of $z^{\prime}-NB921$ vs. $NB921$, 
respectively. Here the matched continuum for NB816, C816, 
is defined as 
$f_{\rm C816} = 0.62 f_{\rm i^\prime} + 0.38 f_{\rm z^\prime}$.
The 3$\sigma$ limiting magnitude of C816 is 27.05 mag, before
Galactic reddening correction. We use the $z^\prime$-band
magnitude as the continuum for NB921. Note that there are 
+0.10 mag and --0.05 mag offsets in the color ($C816-NB816$ 
and $z^{\prime}-NB921$, respectively) distribution of the 
detected objects, as mentioned also by Ly et al. (2007). By 
requiring a minimum NB816 excess of 0.3 mag (i.e., 
$C816-NB816 > 0.3$, that corresponds to 
$EW_{\rm obs} \ga 45 {\rm \AA}$), there are 3 IA598-NB816 
dual-excess objects with an NB816 excess significant at 
higher than 3$\sigma$, and 1 other IA598-NB816 dual-excess 
object with an NB816 excess significance higher than 
2$\sigma$ (Figure 6). Adopting a minimum NB921 excess of 
0.15 mag (i.e., $z^{\prime}-NB921 > 0.15$, that corresponds 
to $EW_{\rm obs} \ga 20 {\rm \AA}$), there are 6 IA679-NB921 
dual-excess objects with a significance of the NB921 excess 
larger than 3$\sigma$. There are no IA679-NB921 dual-excess 
objects with a significance of the NB921 excess between 
2$\sigma$ and 3$\sigma$ (Figure 7).

\subsection{Selection of Ly$\alpha$-He {\sc ii} dual emitters}

In Table 2, we give the photometric properties of the 4 
IA598-NB816 dual-excess objects and of the 6 IA679-NB921 
dual-excess objects. Their spectral energy distributions 
(SEDs) are shown in Figures 8 and 9, respectively. Except 
for a faint IA598-NB816 dual-excess object, IA598\_117217 
(the only object with an NB-excess significance below 
3$\sigma$), the photometric properties of all of the IA-NB 
dual-excess objects are apparently inconsistent with the 
interpretation that they are Ly$\alpha$-He {\sc ii} dual 
emitters at $4.0 \la z \la 4.6$. This is because the IA-NB 
dual-excess objects show relatively blue $B-V$ colors, 
unlike star-forming galaxies at $z \ga 4$, which are 
expected to have B fluxes reduced by Lyman absorption. In 
Figure 10, the predicted $B-V$ colors of galaxy spectral 
models in the observed frame are plotted as functions of 
redshift. We have used the galaxy SED models by 
Bruzual \& Charlot (2003) combined with the cosmic 
transmission by Madau et al. (1996). The galaxy models are 
selected to have solar ($Z = Z_\odot$) or sub-solar 
($Z = 0.005 Z_\odot$) metallicity, a Salpeter IMF, and
an exponentially declining 
star-formation history with $\tau = 1$ Gyr. It is clear 
that galaxies at $z \sim 4.0$ are expected to have 
$B-V \sim 1.8$, or even redder if dust reddening effects 
and the contribution of Ly$\alpha$ photons to the V-band 
flux are taken into account. Galaxies at $z \sim 4.6$ are 
expected to have $B-V >3$, due to the dropout of B-band 
flux caused by redshifted Lyman-limit absorption at 
$912(1+z){\rm \AA} \sim 5100{\rm \AA}$. 

The three IA598-NB816 dual-excess objects (again except 
for IA598\_117217) show $B-V \sim 0.4$, that is consistent 
with the interpretation that they are star-forming
galaxies at $z < 1$. All of the six IA679-NB921 
dual-excess objects show $-0.1 \la B-V \la +0.8$, again 
consistent with the interpretation that those are 
star-forming galaxies at $z < 1$. Note that, based on the 
broad-band classification method for emission-line 
galaxies by Ly et al. (2007), we confirmed that the NB 
excess of the IA-NB dual-excess objects is consistent with 
H$\alpha$ emission or [O~{\sc iii}] emission. We also 
confirmed that all of the 10 IA-NB dual-excess objects are 
detected in the newly obtained U-band data\footnote{
  The U-band data, obtained recently at the Kitt Peak
  National Observatory Mayall 4m telescope, are now in the
  final analysis stages and will be published in a
  forthcoming paper (Ly et al., in preparation).
}, which strongly supports our conclusions that these IA-NB
dual emitters are at $z<1$.

Several IA-NB dual emitters have a large ratio of the 
NB-excess flux to the IA-excess flux, as shown in Table 3. 
The ratios of the inferred EWs, $EW$(IA) to $EW$(NB) 
estimated following the manner of Fujita et al. (2003b), 
can be used to obtain rough estimates of the flux ratios 
of the two emission lines. Ly$\alpha$-He {\sc ii} dual 
emitters (i.e., PopIII-hosting galaxies) cannot have such 
a large flux ratios of He~{\sc ii}/Ly$\alpha$ (which 
should instead be $\la 0.1$, depending on the adopted 
PopIII models; e.g., Schaerer 2003). In contrast, 
star-forming galaxies have the flux ratio of 
[O~{\sc iii}]/[O~{\sc ii}] $\sim$ 0.1--10 (depending on 
the gas metallicity and/or the ionization parameter; e.g., 
Kewley \& Dopita 2002; Nagao et al. 2006) and those of
(H$\alpha$+[N~{\sc ii}])/H$\beta$ $\ga$ 3 (because the 
case B flux ratio of H$\alpha$/H$\beta$ $\sim$ 3; e.g.,
Osterbrock 1989). Therefore, the ratios of the IA excess 
to the NB excess observed in all IA-NB dual-excess 
objects are more consistent with the 
[O~{\sc iii}]/[O~{\sc ii}] or 
(H$\alpha$+[N~{\sc ii}])/H$\beta$ flux ratio of 
star-forming galaxies at $z < 1$, rather than the 
He~{\sc ii}/Ly$\alpha$ of the PopIII-hosting galaxies.

One interesting object is IA598\_117217, that shows only
marginal NB excess with a $\sim 2 \sigma$ significance. 
Its $B-V$ color is $\sim 1.7$ mag, which is consistent 
with the color of galaxies at $z \sim 4.0$. However, this 
object has a significant NB921 excess in addition to 
IA598 and NB816 excess (Figure 8). The corresponding 
rest-frame wavelength is 
$\lambda_{\rm rest} \sim 1840{\rm \AA}$ for $z \sim 4.0$ and 
$\lambda_{\rm rest} \sim 5641{\rm \AA}$ for $z \sim 0.63$,
where no strong emission line is expected in either cases. 
One possibility is that this object harbors an active 
galactic nucleus (AGN). Since the data of different bands 
were obtained in different observing runs that span 6 
years, the time variation of the AGN may cause a spurious 
IA and/or NB excess that is not related to emission lines. 
See, e.g., Morokuma et al. (2008) for the time variation 
of photometric properties of faint AGNs in deep-survey 
data. Although the nature of this object is not clear, we 
do not regard it as a strong candidate of 
Ly$\alpha$-He {\sc ii} dual emission at $z \sim 4.0$.

We thus conclude that most of the 10 IA-NB dual emitters
found in this survey are [O~{\sc ii}]-[O~{\sc iii}] or
H$\beta$-(H$\alpha$+[N~{\sc ii}]) dual emitters, and 
that photometric candidates of Ly$\alpha$-He {\sc ii} 
dual emitters have not been found. Note that the inferred 
$EW_{\rm rest}$ of them are extremely large (Table 3). 
Some objects have $EW_{\rm rest} > 100 {\rm \AA}$. 
Although such large-EW objects at $z < 1$ are very rare, 
some other NB and IA surveys also found such interesting 
objects (Ajiki et al. 2006; Kakazu et al. 2007). 
Kakazu et al. (2007) investigated spectra of NB-selected
emission-line galaxies with $EW_{\rm rest} \ga 100{\rm \AA}$
and found that some of the targets are extremely 
metal-poor galaxies (XMPGs, whose oxygen abundance is
12 + log(O/H) $<$ 7.65; e.g., Kniazev et al. 2003).
Since some of our low-$z$ IA-NB dual emitters have
similar observational properties as XMPGs,
follow-up spectroscopy is of interest.

\section{Discussion}

In the previous section, we showed that there are neither
Ly$\alpha$-He {\sc ii} dual emitters with 
$EW_{\rm obs}$(He~{\sc ii}) $\ga 45{\rm \AA}$ at
$3.93 \la z \la 4.01$, nor those with
$EW_{\rm obs}$(He~{\sc ii}) $\ga 20{\rm \AA}$ at
$4.57 \la z \la 4.65$, detected in the SDF area at our 
limiting magnitudes. Here we discuss the implication of
these results on the abundance of the PopIII-hosting 
galaxies at these redshift ranges. 

Schaerer (2003) investigated the temporal evolution of 
$EW$(He~{\sc ii}) for PopIII stellar clusters by assuming IMFs
with a Salpeter slope, considering the following three cases; 
($M_{\rm low}$, $M_{\rm up}$) = (1$M_\odot$, 100$M_\odot$),
(1$M_\odot$, 500$M_\odot$), and (50$M_\odot$, 500$M_\odot$),
where $M_{\rm low}$ and $M_{\rm up}$ are the lower and
upper mass cut-off values. Here we focus on the predictions
of $EW$(He~{\sc ii}) in the case of 
($M_{\rm low}$, $M_{\rm up}$) = (50$M_\odot$, 500$M_\odot$),
because numerical simulations suggest IMFs biased
toward very high masses (e.g., Bromm et al. 1999; 
Nakamura \& Umemura 2001; Abel et al. 2002; Bromm et al. 
2002). Although the predicted $EW$(He~{\sc ii}) reaches 
$\sim 100 {\rm \AA}$ at the zero age of the PopIII star 
formation, it rapidly decreases and becomes undetectable 
at an age of $\sim 2$ Myr, if an instantaneous burst is 
assumed. In the case of constant star-formation, at the 
equilibrium stage the predicted $EW$(He~{\sc ii}) is 
$\sim 20 {\rm \AA}$. This corresponds to 
$EW_{\rm obs}$(He~{\sc ii}) $\sim 100 {\rm \AA}$ at 
$z \sim 4.0$ and $EW_{\rm obs}$(He~{\sc ii}) 
$\sim 110 {\rm \AA}$ at $z \sim 4.6$ (or equivalently, 
a NB excess of $\sim$0.6 mag for both 
redshift ranges), which are much larger than the detection 
limit of our observation. This suggests that PopIII-hosting 
galaxies younger than $\sim 2$ Myr or having on-going 
PopIII-formation, with an enough high star-formation rate 
($SFR_{\rm PopIII}$), should be detected in our surveys.

In the following we quantify the minimum $SFR_{\rm PopIII}$
to which our survey is sensitive. By assuming constant star 
formation, the He~{\sc ii} luminosity can be written as
follows:
\begin{equation}
    L({\rm He II}) = 
        ( 1 - f_{\rm esc} ) f_{1640}
        \left( \frac{SFR_{\rm PopIII}}{M_\odot {\rm yr}^{-1}} \right),
\end{equation}
where $f_{\rm esc}$ is the escape fraction of the 
He$^+$-ionizing photon and $f_{1640}$ is a proportionality
constant. In the following we assume that $f_{\rm esc}$ is 
negligibly small. Schaerer (2003) predicted 
$f_{1640} = 6.01 \times 10^{41}$ for PopIII stellar clusters with 
($M_{\rm low}$, $M_{\rm up}$) = (50$M_\odot$, 500$M_\odot$). 
This corresponds to $F$(He~{\sc ii}) = $3.91 \times 10^{-18}$ 
($SFR_{\rm PopIII}/M_\odot$ yr$^{-1}$) for $z = 4.0$ and 
$F$(He~{\sc ii}) = $2.81 \times 10^{-18}$ 
($SFR_{\rm PopIII}/M_\odot$ yr$^{-1}$) for $z = 4.6$, in 
units of erg s$^{-1}$ cm$^{-2}$. Since the 3$\sigma$ limiting 
flux of objects with $EW_{\rm obs} = 100{\rm \AA}$ in our 
NB816 observation is $7.4 \times 10^{-18}$ erg s$^{-1}$ 
cm$^{-2}$ and that of objects with 
$EW_{\rm obs} = 110{\rm \AA}$ in our NB921 observation is 
$5.9 \times 10^{-18}$ erg s$^{-1}$ cm$^{-2}$, our survey can 
detect PopIII-hosting galaxies if their SFR is higher than 
$\sim 2 M_\odot$ yr$^{-1}$. Therefore, the non-detection of 
Ly$\alpha$-He~{\sc ii} dual emitters suggests that there are 
no PopIII-hosting galaxies with 
$SFR_{\rm PopIII} \ga 2 M_\odot$ yr$^{-1}$ at 
$4.0 \la z \la 4.6$ toward the SDF, in a volume of 
$4.03 \times 10^5$ Mpc$^3$. This result implies an 
upper-limit of the PopIII SFR density of
$SFRD_{\rm PopIII} < 5 \times 10^{-6} M_\odot$ yr$^{-1}$
Mpc$^{-3}$, if taking only galaxies with 
$SFR_{\rm PopIII} > 2 M_\odot$ yr$^{-1}$ into account.
Note that the inferred upper limit on $SFR_{\rm PopIII}$ is
uncertain, since the predicted flux of He~{\sc ii} for a
given $SFR_{\rm PopIII}$ strongly depends on the assumed IMF 
(e.g., Schaerer 2003). It also depends on the evolutionary
processes of PopIII stars, especially the mass loss during 
their evolution (e.g., El Eid et al. 1983; Tumlinson et al. 
2001; Schaerer 2002).

Some theoretical studies suggest that the volume-averaged 
IGM metallicity quickly reached $Z_{\rm crit} = 10^{-4} Z_\odot$ 
at $z>10$ (e.g., Mackey et al. 2003; Yoshida et al. 2004; 
Tornatore et al. 2007), where $Z_{\rm crit}$ is the critical 
metallicity, below which very massive stars could be formed.
However, this does not necessarily suggest that the 
formation of PopIII stars was terminated at such a high 
redshift, because of the inhomogeneous metal distribution in 
the early universe (e.g., Scannapieco et al. 2003, 2006;
Brook et al. 2007; Tornatore et al. 2007). As demonstrated by 
Scannapieco et al. (2003), the redshift evolution of the 
$SFR_{\rm PopIII}$ density in the universe depends 
sensitively on some PopIII model parameters, especially the 
feedback efficiency that is closely related to the PopIII 
IMF (see also Scannapieco et al. 2006). Low-feedback models 
of Scannapieco et al. (2003) predict a large fraction 
($\sim$ 30\%) of PopIII-hosting galaxies among LAEs at 
$4.0 \la z \la 4.6$ with log 
$L$(Ly$\alpha$) $\sim 10^{43}$ ergs s$^{-1}$ (again 
$M_{\rm low} = 50 M_\odot$ is assumed here). A similarly 
large fraction of PopIII-hosting galaxies among high-$z$ 
LAEs is also inferred by Dijkstra \& Wyithe (2007). Since 
the number density of LAEs with this luminosity at similar 
redshifts is $\sim 10^{-5} - 10^{-4}$ Mpc$^{-3}$ 
(Ouchi et al. 2003; Yamada et al. 2005; Ouchi et al. 2008), 
the number of PopIII-hosting galaxies in our survey, 
expected by such low-feedback models, is roughly 1 to 10. 
Therefore, the non-detection in our ``Ly$\alpha$-He~{\sc ii} 
dual emitters'' survey may suggest that low-feedback models 
are not appropriate, and that PopIII stars may instead by 
characterized by a relatively large feedback efficiency.

This photometric survey for Ly$\alpha$-He {\sc ii} dual 
emitters demonstrated that wide and deep imaging 
observations, combining narrow-band and/or 
intermediate-band filters, are potentially a powerful tool to 
search or constrain the properties of PopIII-hosting galaxies
at high redshifts. The data recently obtained by sensitive 
narrow-band near-infrared surveys at 1.19$\mu$m 
(Willis \& Courbin 2005; Cuby et al. 2007; Willis et al. 2007) 
and similar wide and deep surveys planned in future may be 
useful to search for Ly$\alpha$-He {\sc ii} dual emitters at 
$z \sim 6.3$, by adding data of narrow- or intermediate-band 
observations at $\lambda \sim 8820{\rm \AA}$ to check 
strong Ly$\alpha$ emission. Such a survey is promising,
since $SFR_{\rm PopIII}$ increases at higher redshifts (see, 
e.g., Dijkstra \& Wyithe 2007). In future observational 
searches for Ly$\alpha$-He {\sc ii} dual emitters, serious 
sources of contamination would be [O~{\sc ii}]-[O~{\sc iii}]
and H$\beta$-(H$\alpha$+[N~{\sc ii}]) dual emitters, as 
demonstrated in this paper. In addition to broad-band color 
criteria, the flux (or EW) ratio of the dual excesses is 
also a powerful diagnostic to discriminate the populations 
and to identify Ly$\alpha$-He {\sc ii} dual emitters among 
the photometric candidates.

\acknowledgments

  We thank A. Marconi, K. Shimasaku, M. Morokuma, T., Yagi, K. 
  Omukai, N. Yoshida, and the anonymous referee for their 
  useful comments, and Y. Ideue, S. Mihara and A. Nakajima for 
  assisting the Subaru observation runs. We also thank K. Ota 
  for providing us an optical sky spectrum. This research is 
  based on data collected at the Subaru Telescope, which is 
  operated by the National Astronomical Observatory of Japan. 
  We are grateful to all members of the Subaru Deep Field 
  project and to the staff of the Subaru Telescope, especially 
  to the Subaru support astronomers, H. Furusawa and M. Takami. 
  We wish to recognize and acknowledge the very significant
  cultural role and relevance that the summit of Mauna Kea 
  has always had within the indigenous Hawaiian community.
  We are most fortunate to have the opportunity to conduct
  observations from this mountain.
  TN and SSS are financially supported by the Japan Society 
  for the Promotion of Science (JSPS) through JSPS Research 
  Fellowship for Young Scientists.

\clearpage

\begin{figure}
\epsscale{1.0}
\includegraphics[width=9.0cm,angle=-90]{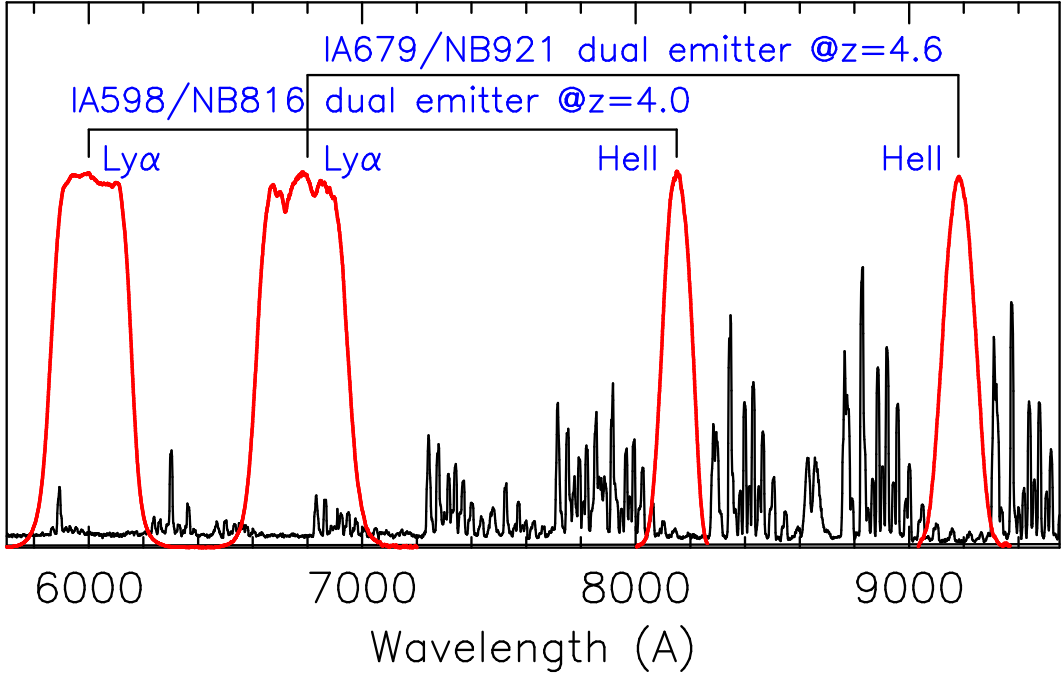}
\caption{
   Schematic view of the selection of Ly$\alpha$-He {\sc ii}
   dual emitters. The black solid spectrum denotes a typical
   background sky spectrum. Red solid curves denote the
   filter transmission curves of IA598, IA679, NB816, and
   NB921. The dual excess of the combination of IA598 and
   NB816, and that of IA679 and NB921 correspond to
   $z \sim 4.0$ and $z \sim 4.6$, respectively.
\label{fig1}}
\end{figure}

\begin{figure}
\epsscale{1.0}
\includegraphics[width=7.5cm,angle=-90]{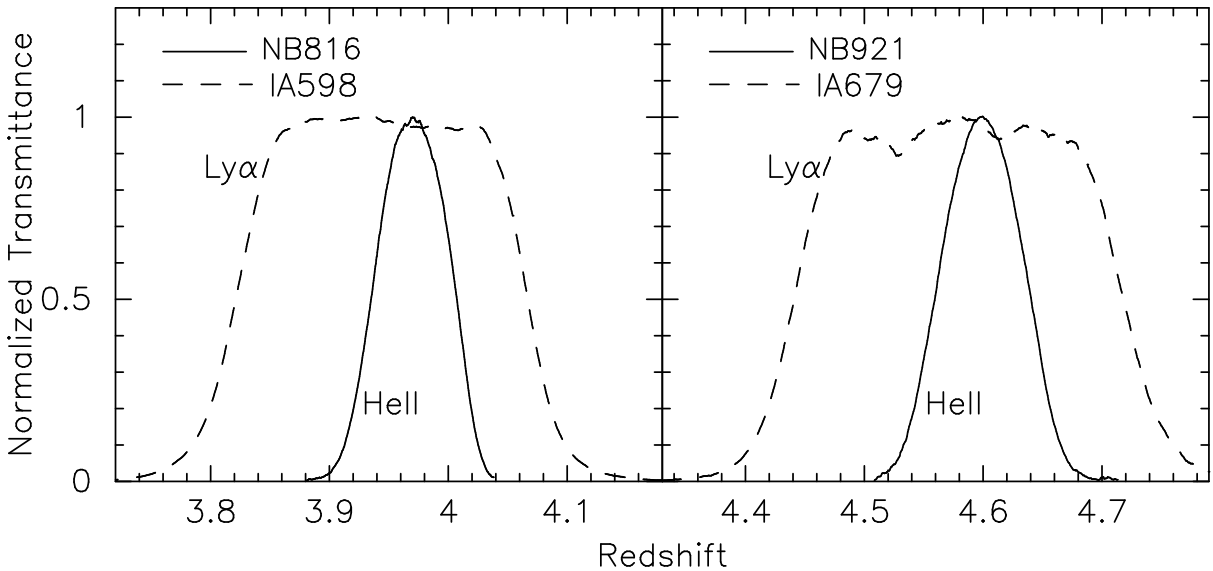}
\caption{
   Transmission curves of IA598 and NB816 (dashed and solid 
   lines in the left panel), and those of IA679 and NB921 
   (dashed and solid lines in the right panel) normalized by
   their peak transmittances are shown as a function of
   redshift.
\label{fig2}}
\end{figure}

\begin{figure}
\epsscale{1.0}
\includegraphics[width=15.5cm,angle=-90]{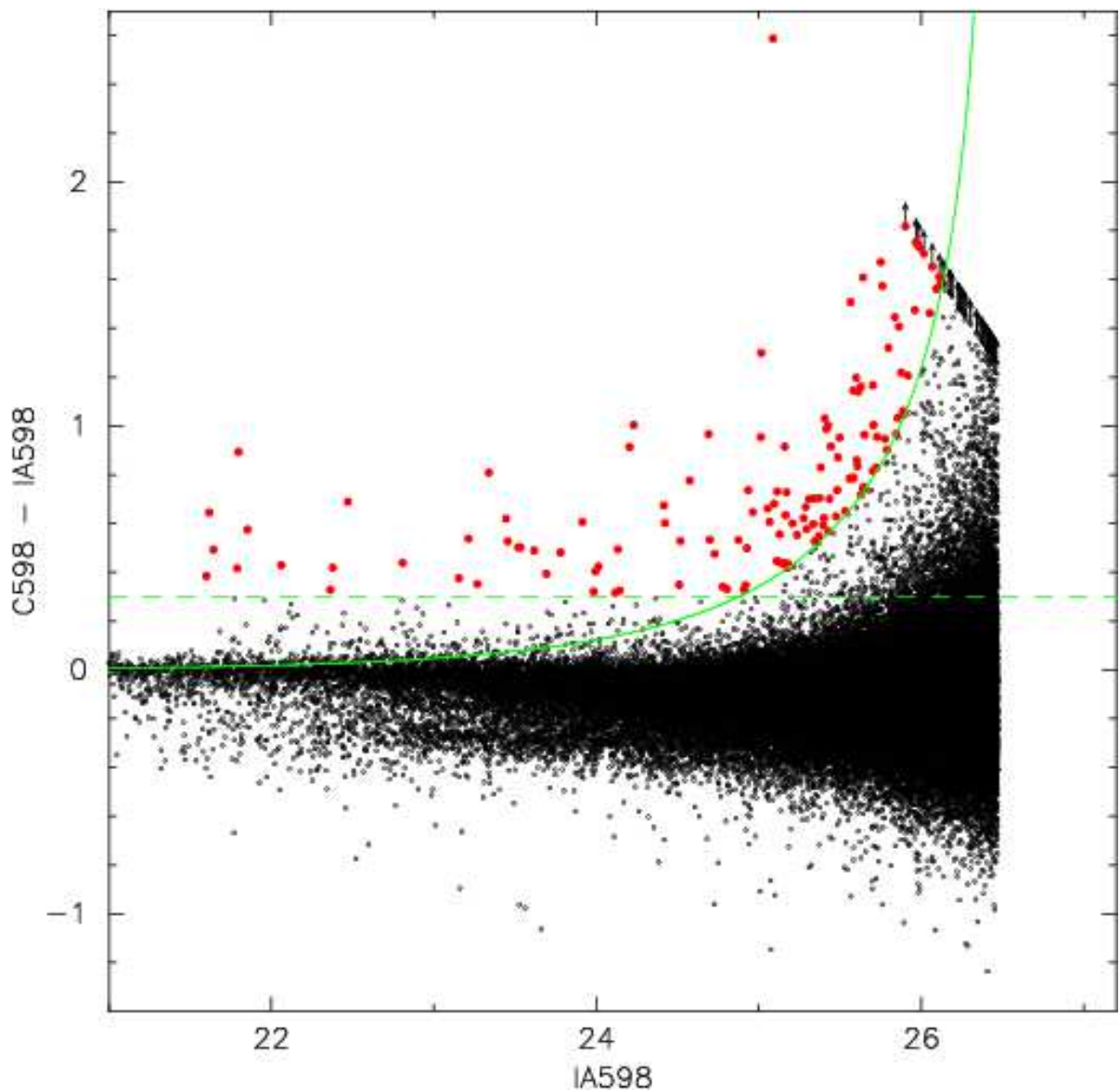}
\caption{
   IA598-detected objects (which are also detected at 2$\sigma$
   in both $R_{\rm C}$ and $i^\prime$ images), plotted on a 
   diagram of $C598$--$IA598$ vs $IA598$ (black dots; 63660 objects). 
   The dashed green horizontal line denotes the excess criterion 
   limit, $C598$--$IA598=0.3$. The solid green curve denotes 
   3$\sigma$ uncertainty in the C598--IA598 color. Objects with 
   a lower-limit on the color of C598--IA598 are shown with 
   arrows. Red, filled circles denote the selected IA598-excess 
   galaxies (133 objects). 
\label{fig3}}
\end{figure}

\begin{figure}
\epsscale{1.0}
\includegraphics[width=15.5cm,angle=-90]{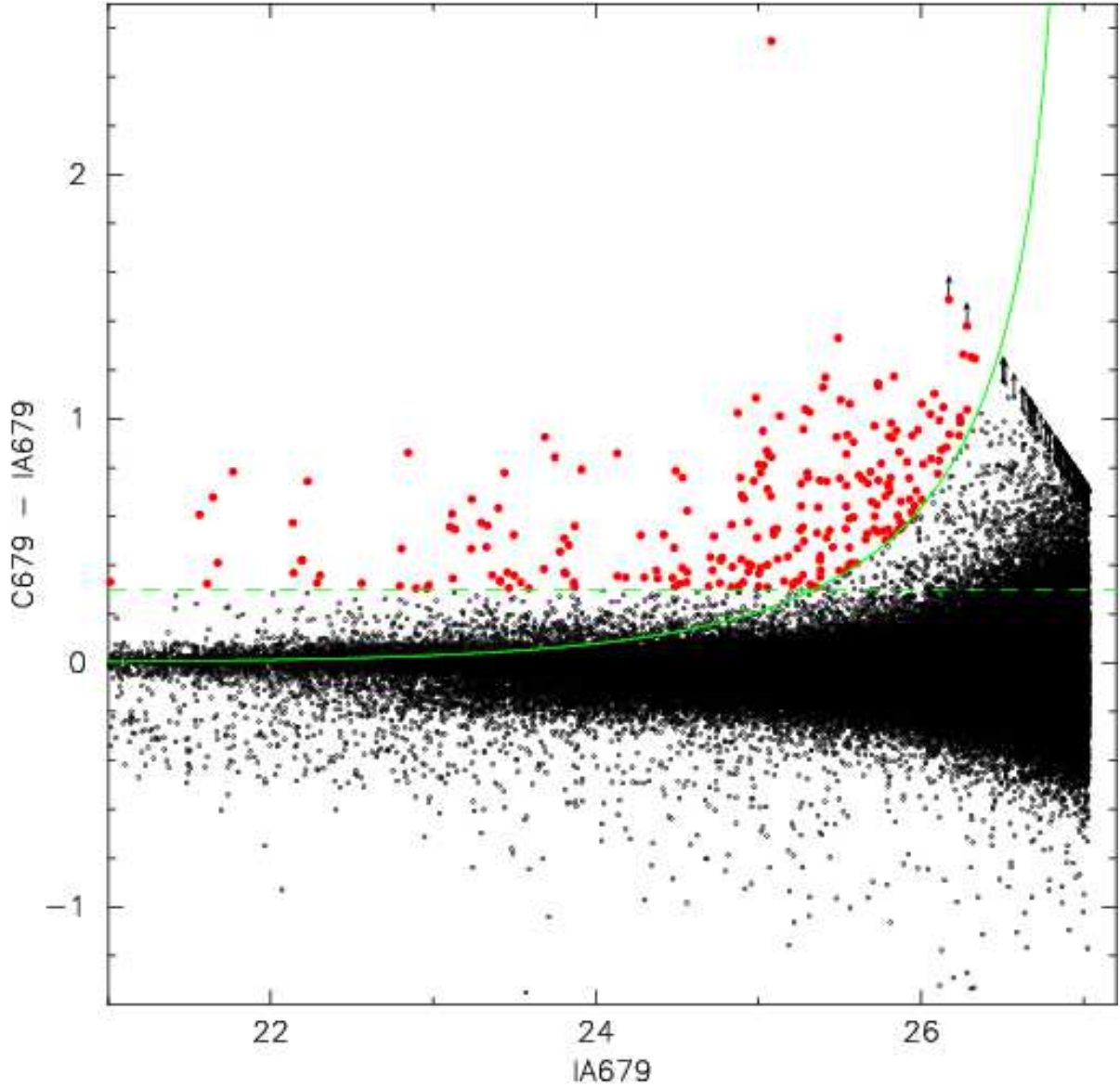}
\caption{
   IA679-detected objects (which are also detected at 2$\sigma$
   in both $R_{\rm C}$ and $i^\prime$ images), plotted on a 
   diagram of $C679$--$IA679$ vs $IA679$ (black dots; 97234 objects). 
   The dashed green horizontal line denotes the excess criterion 
   limit, $C679$--$IA679=0.3$. The solid green curve denotes 
   3$\sigma$ uncertainty in the C679--IA679 color. Objects with
   a lower-limit on the color of C679--IA679 are shown with 
   arrows. Red, filled circles denote the selected IA679-excess 
   galaxies (234 objects). 
\label{fig4}}
\end{figure}

\begin{figure}
\epsscale{1.0}
\includegraphics[width=15.5cm,angle=-90]{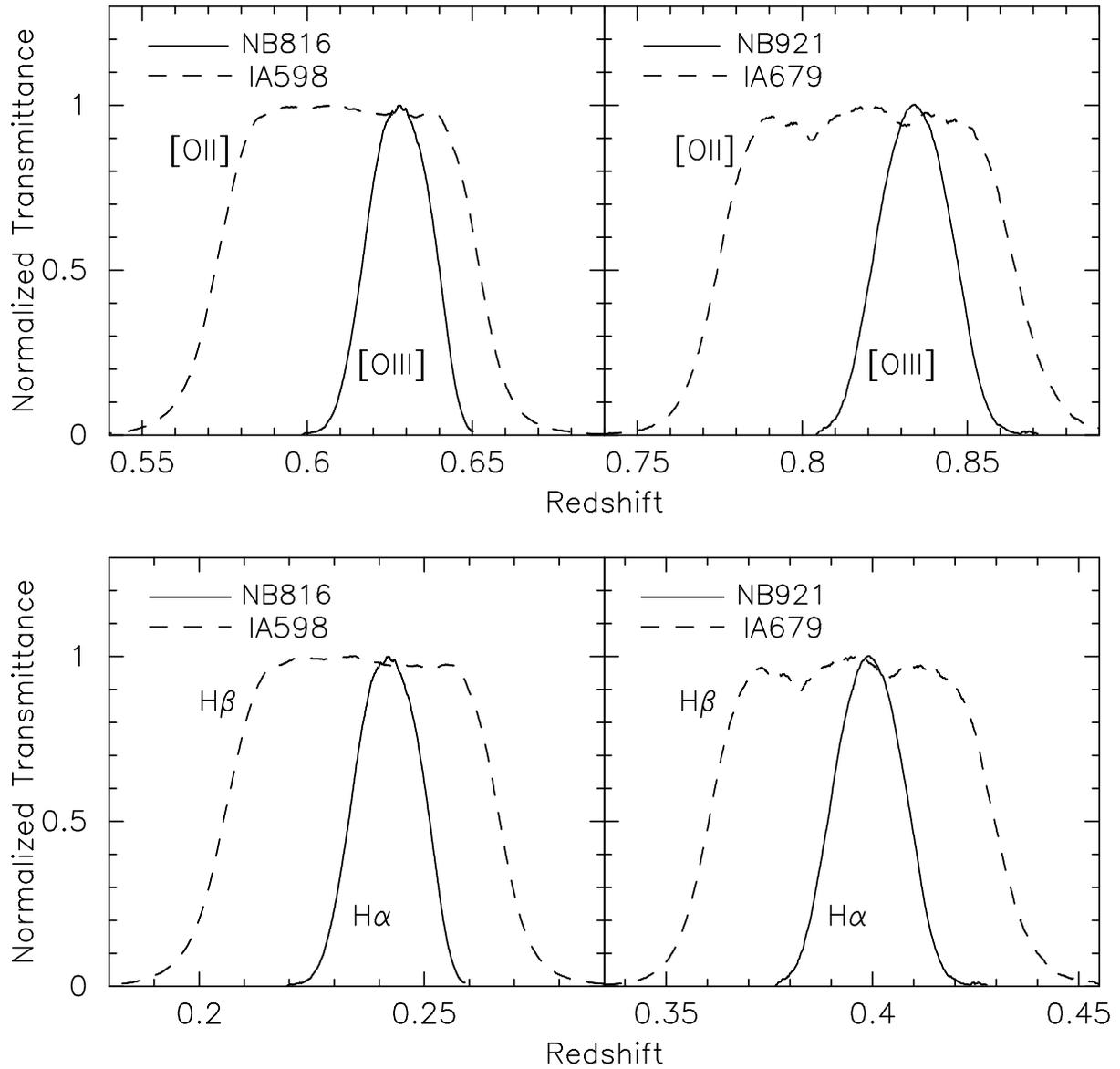}
\caption{
   Same as Figure 2 but for [O~{\sc ii}]-[O~{\sc iii}] 
   dual emitters (upper panels) and for 
   H$\beta$-(H$\alpha$+[N~{\sc ii}]) dual emitters
   (lower panels).
\label{fig5}}
\end{figure}

\begin{figure}
\epsscale{1.0}
\includegraphics[width=15.5cm,angle=-90]{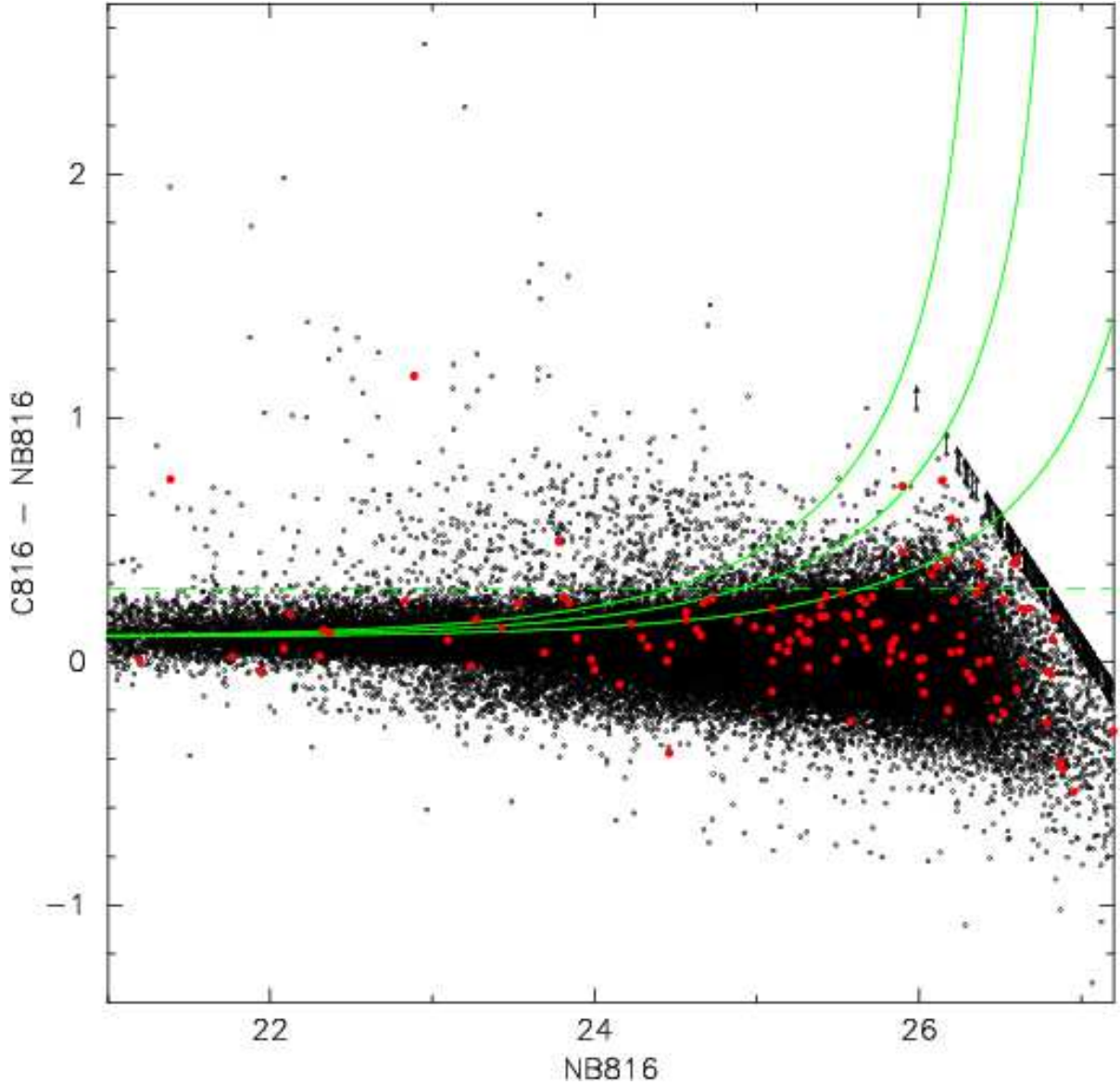}
\caption{
   IA598-detected objects (which are also detected at 2$\sigma$
   in both $R_{\rm C}$ and $i^\prime$ images), plotted on a diagram
   of $C816-NB816$ vs $NB816$ (black dots; 63660 objects). The
   dashed green horizontal line denotes the excess criterion limit, 
   $C816-NB816=0.3$. The solid green curve denotes 3$\sigma$, 
   2$\sigma$, and 1$\sigma$ uncertainty in the C816--NB816 
   color, taking a +0.10 mag offset into account (see text). Objects 
   with a lower-limit on the color of C816--IA816 are shown with 
   arrows. Red, filled circles denote the selected IA598-excess 
   galaxies (133 objects). 
\label{fig7}}
\end{figure}

\begin{figure}
\epsscale{1.0}
\includegraphics[width=15.5cm,angle=-90]{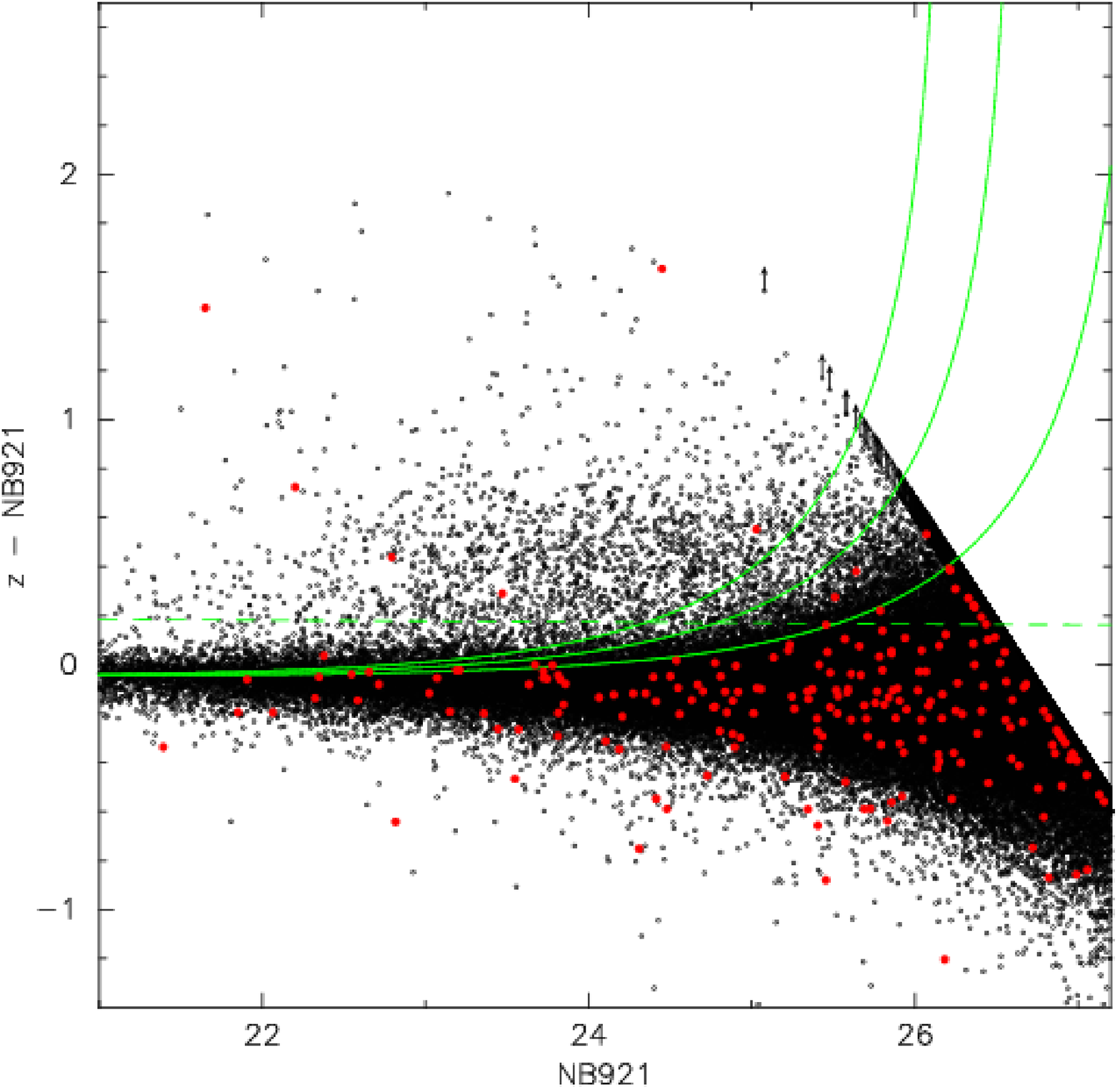}
\caption{
   IA679-detected objects (which are also detected at 2$\sigma$
   in both $R_{\rm C}$ and $i^\prime$ images), plotted on a diagram
   of $z^{\prime}-NB921$ vs $NB921$ (black dots; 97234 objects). 
   The dashed green horizontal line denotes the excess criterion 
   limit, $z^{\prime}-NB921=0.3$. The solid green curve denotes 
   3$\sigma$, 2$\sigma$, and 1$\sigma$ uncertainty in the 
   $z^{\prime}$--NB921 color, taking a --0.05 mag offset into 
   account (see text). Objects with a lower-limit on the color of 
   $z^{\prime}$--NB921 are shown with arrows. Red, filled circles 
   denote the selected IA679-excess galaxies (234 objects). 
\label{fig7}}
\end{figure}

\begin{figure}
\epsscale{1.0}
\includegraphics[width=14.0cm,angle=-90]{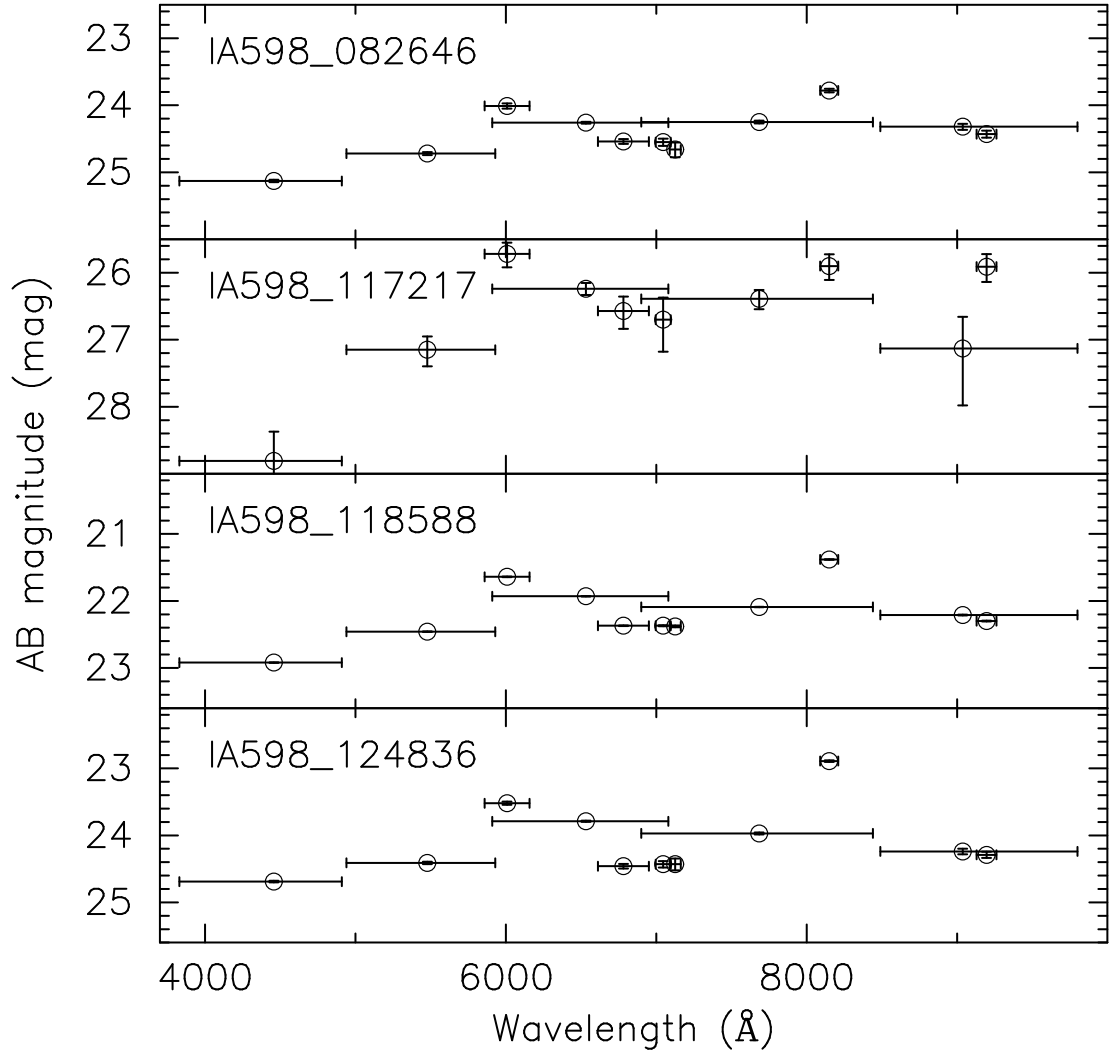}
\caption{
   SEDs of the IA598-NB816 dual emitters. Error bars in the
   y-axis direction denote the 1$\sigma$ photometric errors.
   The ID of each object is shown at the upper-left corner of 
   each panel.
\label{fig8}}
\end{figure}

\begin{figure}
\epsscale{1.0}
\includegraphics[width=20.0cm,angle=-90]{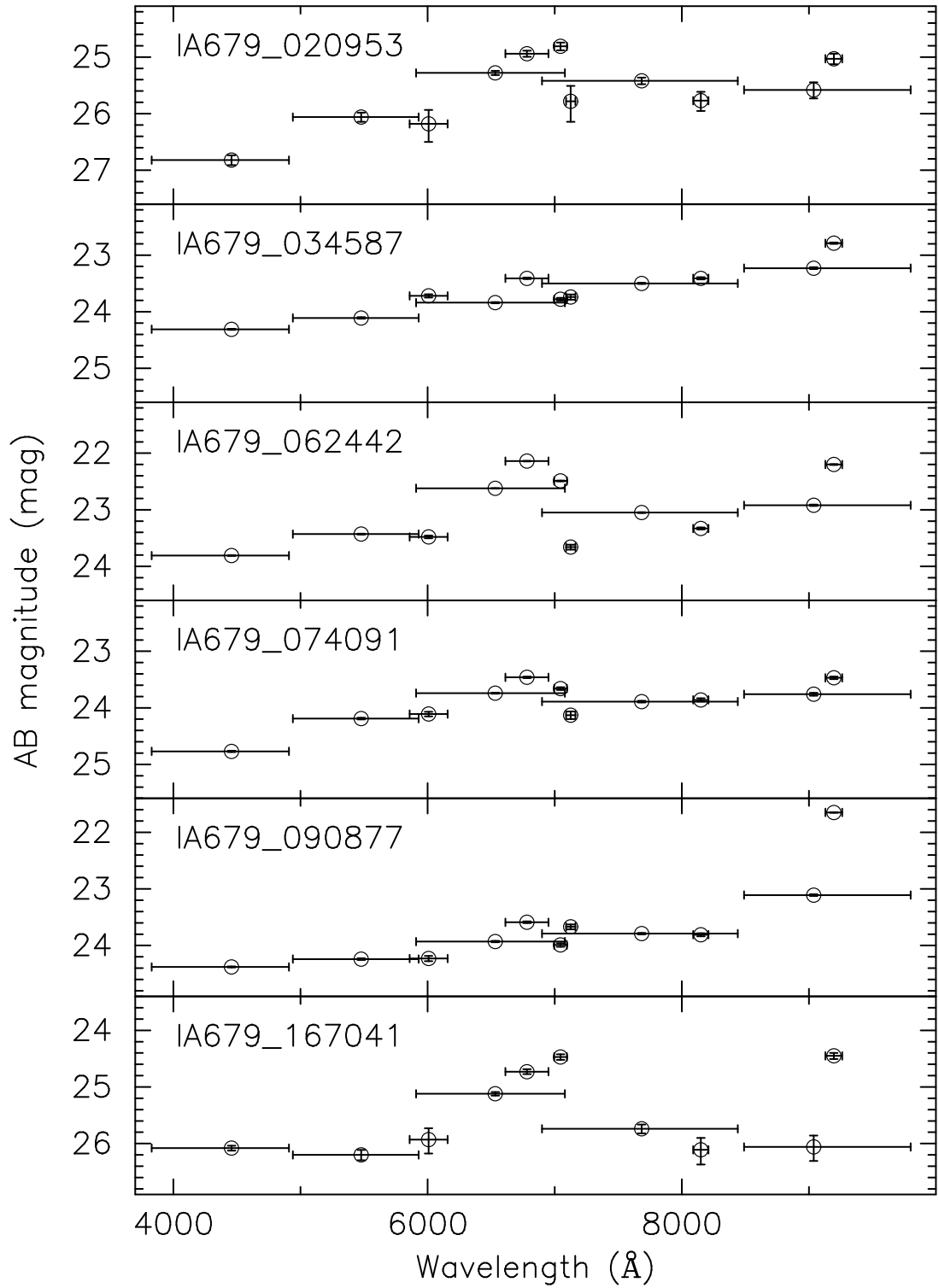}
\caption{
   Same as Figure 8 but for the IA679-NB921 dual emitters.
\label{fig9}}
\end{figure}

\begin{figure}
\epsscale{1.0}
\includegraphics[width=20.0cm,angle=-90]{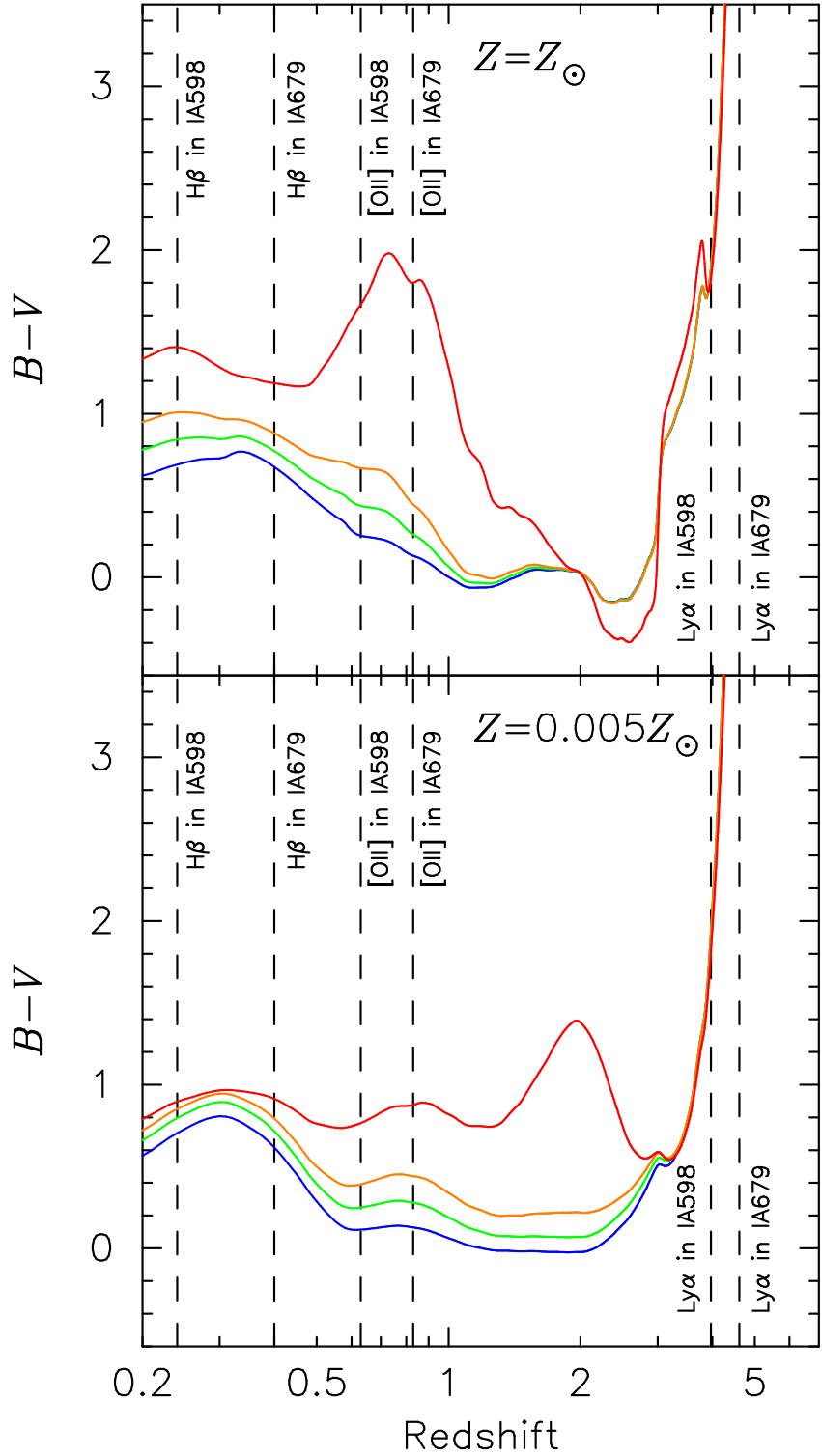}
\caption{
   Predicted $B-V$ color of galaxy models (Bruzual \& Charlot 
   2003) as a function of redshift. The galaxy models have 
   Salpeter IMF and an exponentially declining star-formation 
   history with $\tau = 1$ Gyr. Blue, green, yellow, and red solid 
   lines corresponds to the model predictions for an age of
   2, 3, 4, and 8 Gyr, respectively. Vertical dashed lines denote 
   the redshifts where the IA-NB dual-excess objects are 
   expected to be located. Models with a metallicity of 
   $Z=Z_\odot$ and with that of $Z=0.005Z_\odot$ are shown 
   in the upper and lower panels, respectively.
\label{fig10}}
\end{figure}

\clearpage

\begin{deluxetable}{cllcl}
\tablewidth{0pt}
\tablecaption{Available optical photometric data in the SDF}
\tablehead{
  \colhead{Filter                    } &
  \colhead{$t_{\rm exp}$             } &
  \colhead{$m_{\rm lim}$$^\mathrm{a}$} &
  \colhead{$A_V$$^\mathrm{b}$        } &
  \colhead{Ref.$^\mathrm{c}$         }\\
  \colhead{                          } &
  \colhead{(min.)                    } &
  \colhead{(mag)                     } &
  \colhead{(mag)                     } &
  \colhead{                          }
}
\startdata
$B$         & 595 & 28.45 & 0.07 & 1 \\
$V$         & 340 & 27.74 & 0.05 & 1 \\
$R_{\rm C}$ & 600 & 27.80 & 0.04 & 1 \\
$i^\prime$  & 801 & 27.43 & 0.03 & 1 \\
$z^\prime$  & 504 & 26.62 & 0.02 & 1 \\
IA598       & 111 & 26.52 & 0.05 & 2 \\
IA679       & 231 & 27.07 & 0.04 & 2 \\
NB704       & 198 & 26.67 & 0.04 & 3, 4 \\
NB711       & 162 & 25.99 & 0.04 & 4, 5, 6        \\
NB816       & 600 & 26.63 & 0.03 & 1, 4, 7        \\
NB921       & 899 & 26.54 & 0.02 & 1, 4, 8, 9, 10 \\
NB973       & 900 & 25.5  & 0.02 & 11, 12         \\
\enddata
\tablenotetext{a}{
  3$\sigma$ limiting magnitude for 2$^{\prime\prime}\phi$
  apertures, without the Galactic reddening correction.}
\tablenotetext{b}{
  Galactic extinction for each band (Schlegel et al. 1998), 
  calculated by adopting the Galactic extinction curve 
  (Cardelli et al. 1989).}
\tablenotetext{c}{
  References:
   (1) Kashikawa et al. (2004).
   (2) This work.
   (3) Shimasaku et al. (2004).
   (4) Ly et al. (2007).
   (5) Ouchi et al. (2003).
   (6) Shimasaku et al. (2003).
   (7) Shimasaku et al. (2006).
   (8) Kodaira et al. (2003).
   (9) Taniguchi et al. (2005a).
   (10) Kashikawa et al. (2006).
   (11) Iye et al. (2006).
   (12) Ota et al. (2007).}
\end{deluxetable}

\clearpage

\begin{deluxetable}{cccccccccccc}
%\rotate
\tabletypesize{\footnotesize}
\tablewidth{0pt}
\tablecaption{Photometric properties of IA-NB dual emitters$^\mathrm{a}$}
\tablehead{
  \colhead{ID         }      &
  \colhead{$B$        }      &
  \colhead{$V$        }      &
  \colhead{$R_{\rm C}$}      &
  \colhead{$i^\prime$ }      &
  \colhead{$z^\prime$ }      &
  \colhead{IA598}  &
  \colhead{IA679}  &
  \colhead{NB704}  &
  \colhead{NB711}  &
  \colhead{NB816}  &
  \colhead{NB921}  \\
  \colhead{     }            &
  \colhead{(mag)}            &
  \colhead{(mag)}            &
  \colhead{(mag)}            &
  \colhead{(mag)}            &
  \colhead{(mag)}            &
  \colhead{(mag)}            &
  \colhead{(mag)}            &
  \colhead{(mag)}            &
  \colhead{(mag)}            &
  \colhead{(mag)}            &
  \colhead{(mag)} 
}
\startdata
IA598\_082646 & 25.13 & 24.72 & 24.26 & 24.25 & 24.32 & 
                24.01 & 24.54 & 24.55 & 24.66 & 23.78 & 24.43 \\
IA598\_117217 & 28.81 & 27.15 & 26.24 & 26.39 & 27.13 & 
                25.72 & 26.57 & 26.70 &  ---  & 25.90 & 25.91 \\
IA598\_118588 & 22.92 & 22.46 & 21.93 & 22.09 & 22.21 & 
                21.64 & 22.37 & 22.37 & 22.38 & 21.38 & 22.30 \\
IA598\_124836 & 24.69 & 24.41 & 23.79 & 23.97 & 24.24 & 
                23.52 & 24.46 & 24.43 & 24.43 & 22.89 & 24.29 \\
IA679\_020953 & 26.82 & 26.06 & 25.28 & 25.42 & 25.58 & 
                26.18 & 24.94 & 24.81 & 25.78 & 25.77 & 25.03 \\
IA679\_034587 & 24.31 & 24.11 & 23.84 & 23.50 & 23.23 & 
                23.72 & 23.41 & 23.78 & 23.74 & 23.41 & 22.79 \\
IA679\_062442 & 23.81 & 23.43 & 22.62 & 23.05 & 22.92 & 
                23.48 & 22.14 & 22.49 & 23.66 & 23.33 & 22.20 \\
IA679\_074091 & 24.77 & 24.19 & 23.74 & 23.89 & 23.76 & 
                24.11 & 23.46 & 23.66 & 24.13 & 23.86 & 23.47 \\
IA679\_090877 & 24.38 & 24.24 & 23.93 & 23.79 & 23.11 & 
                24.23 & 23.59 & 23.99 & 23.67 & 23.81 & 21.65 \\
IA679\_167041 & 26.08 & 26.20 & 25.12 & 25.74 & 26.06 & 
                25.93 & 24.73 & 24.47 &  ---  & 26.11 & 24.45 \\
\enddata
\tablenotetext{a}{
  Corrected for the Galactic extinction.}
\end{deluxetable}

\begin{deluxetable}{ccc}
\tablewidth{0pt}
\tablecaption{Equivalent widths of IA-NB dual emitters}
\tablehead{
  \colhead{ID                } &
  \colhead{$EW_{\rm obs}$(IA)} & 
  \colhead{$EW_{\rm obs}$(NB)}\\
  \colhead{                  } &
  \colhead{(${\rm \AA}$)     } &
  \colhead{(${\rm \AA}$)     }
}
\startdata
IA598\_082646 & 181 $\pm$  20 &  83 $\pm$  8 \\
IA598\_117217 & 483 $\pm$ 157 & 137 $\pm$ 67 \\
IA598\_118588 & 218 $\pm$   2 & 147 $\pm$  1 \\
IA598\_124836 & 224 $\pm$  14 & 299 $\pm$  6 \\
IA679\_020953 & 191 $\pm$  37 &  96 $\pm$ 28 \\
IA679\_034587 & 167 $\pm$   9 &  72 $\pm$  3 \\
IA679\_062442 & 349 $\pm$   4 & 141 $\pm$  2 \\
IA679\_074091 & 154 $\pm$   9 &  43 $\pm$  5 \\
IA679\_090877 & 147 $\pm$  10 & 542 $\pm$  4 \\
IA679\_167041 & 297 $\pm$  36 & 709 $\pm$ 62 \\
\enddata
\end{deluxetable}

\end{document}